\documentclass[a4paper,fleqn,usenatbib]{mnras}

\usepackage{mathptmx}

\usepackage[T1]{fontenc}
\usepackage{ae,aecompl}

\usepackage{enumerate}
\usepackage{graphicx}	
\usepackage{amsmath}	
\usepackage{amssymb}	
\usepackage{longtable,lscape}

\bibpunct{(}{)}{;}{a}{}{,}
\title[]{A massive open cluster hiding in full sight}

\author[I. Negueruela et al.]{
I. Negueruela,$^{1,2}$\thanks{E-mail: ignacio.negueruela@ua.es}
A.-N. Chen\'e,$^{3,4}$
H.~M. Tabernero,$^{2,5}$
R. Dorda,$^{2,6,7}$\newauthor
J.~Borissova,$^{3,8}$
A. Marco,$^{2}$
and 
R.~Kurtev$^{3,8}$\\
$^{1}$Departamento de F\'{\i}sica Aplicada, Facultad de Ciencias, Universidad de Alicante,\\ Carretera de San Vicente s/n, E03690, San Vicente del Raspeig, Spain\\
$^{2}$Departamento de F\'{\i}sica, Ingenier\'{\i}a de Sistemas y
Teor\'{\i}a de la Se\~{n}al, Universidad de Alicante,\\ Carretera de San Vicente s/n,
E03690, San Vicente del Raspeig, Spain\\
$^{3}$Instituto de F\'isica y Astronom\'ia, Universidad de Valpara\'iso, Av. Gran Breta\~na 1111, Playa Ancha, Casilla 5030, Chile.\\
$^{4}$Gemini Observatory, Northern Operations Center, 670 A'ohoku Place, Hilo, HI 96720, USA\\
$^{5}$Instituto de Astrof{\'i}sica e Ci{\^e}ncias do Espa\c{c}o, Universidade do Porto, CAUP, Rua das Estrelas, 4150-762 Porto, Portugal\\
 $^{6}$Instituto de Astrof\'{\i}sica de Canarias,  V\'{\i}a L\'actea s/n, E38200, La Laguna, Tenerife, Spain\\
 $^{7}$Universidad de La Laguna, Dpto. Astrofísica, E-38206 La Laguna, Tenerife, Spain\\
$^{8}$Instituto Milenio de Astrofísica, Nuncio Monse\~{n}or Sotero Sanz 100, Of. 104, Providencia, Santiago, Chile \\
}

\date{Accepted XXX. Received YYY; in original form ZZZ}

\pubyear{2016}

\begin{document}
\label{firstpage}
\pagerange{\pageref{firstpage}--\pageref{lastpage}}
\maketitle

\begin{abstract}
Obscuration and confusion conspire to limit our knowledge of the inner Milky Way. Even at moderate distances, the identification of stellar systems becomes compounded by the extremely high density of background sources. Here we provide a very revealing example of these complications by unveiling a large, massive, young cluster in the Sagittarius arm that has escaped detection until now despite containing more than 30 stars brighter than $G=13$. By combining \textit{Gaia} DR2 astrometry, \textit{Gaia} and 2MASS photometry and optical spectroscopy, we find that the new cluster, which we name Valparaiso~1, located at $\sim2.3\:$kpc, is about 75~Ma old and includes a large complement of evolved stars, among which we highlight the 4~d classical Cepheid CM~Sct and an M-type giant that probably represents the first detection of an AGB star in a Galactic young open cluster. Although strong differential reddening renders accurate parameter determination unfeasible with the current dataset, direct comparison to clusters of similar age suggests that Valparaiso~1 was born as one of the most massive clusters in the Solar Neighbourhood, with an initial mass close to $10^{4}\:\mathrm{M}_{\sun}$.
\end{abstract}




\section{Introduction}

Star formation is a complex hierarchical process. Its outcome depends on many factors that we do not know yet. Giant molecular clouds can give rise to dispersed OB associations that contain no populous clusters, as in Cyg~OB2 \citep{wright16}, but also produce massive, compact clusters, with masses $M_{\mathrm{cl}}\ga10^{4}\:$M$_{\sun}$, such as NGC~3603 \citep[e.g.][]{rochau10}.
Despite this diversity, the mass distribution of open clusters in non-interacting galaxies seems to be quite similar and independent of their star formation rates \citep{fall12}, especially among spirals \citep[e.g.][]{larsen09}. Based on the expectations of such a standard distribution, the Milky way should produce one bound cluster more massive than 10$^4\:$M$_{\sun}$ about every $\sim$500\,000 years \citep{larsen09}. Indeed, about a dozen star clusters with ages $\la$20$\:$Ma and masses above this limit have now been found in our Galaxy \citep{pz10,negueruela14}. There is, however, a general lack of clusters with comparable masses at slightly older ages. In fact, only M11 \citep{santos05, cantat14} has been confirmed to be above the 10$^4\:$M$_{\sun}$ landmark among moderately young clusters. \citet{Bav16}, based on the properties of a eclipsing binary in the cluster, give a turn-off mass of 3.6\:M$_{\sun}$ and thus an age around 250\,Ma, which implies an initial mass well above this value \citep[cf.][]{cantat14}. The heavily obscured cluster Mercer~13 \citep{messineo09} may have a similar age and an equivalent population, but its parameters are still uncertain.

Similar clusters must have been forming throughout the history of the Milky Way. In fact, any old open cluster that has managed to survive to this day in the Galactic gravitational potential must have started as a rather massive object. For instance, \citet{hurley05} estimate that M~67, which is $\sim4$~Ga old, started its life with almost $20\,000\:$M$_{\sun}$ in order to retain $\sim1\,500\:$M$_{\sun}$ at present.
Several authors have proposed the heavily reddened compact cluster GLIMPSE-C01 to be the intermediate-age descendent of a very massive young open cluster. Data are compatible with a 2~Ga cluster still having $\sim 10^5\:$M$_{\sun}$, but also with an old globular cluster that happens to be crossing the Galactic Plane \citep{hare18}. At intermediate ages, the outer Galaxy contains several clusters that seem to have had very high initial masses. Among them, we can count NGC~2477, for which \citet{eigenbrod04} estimate a current mass of $\sim5\,400\:$M$_{\sun}$ in stars more massive than the Sun at an age of 1.0~Ga, which seems confirmed by more recent works. NGC~1817 is similar \citep{mermi03}, while  NGC~2099 is about half this age and at least as massive \citep{kalirai01}. 

Towards the inner Galaxy, there are very few equivalent clusters. NGC~4337, which is some distance above the Plane ($b=+4\fd6$), has an estimated mass $\ga 2 \times 10^{3}\:$M$_{\sun}$ at 1.6~Ga and may be the remnant of a rather massive cluster \citep{seleznev17}, but very few similar objects have been found and studied. Considering that most of the Galactic mass lies towards the inside and that most young massive clusters known are within the Solar circle, there are two immediate explanations\footnote{Many works indicate a very rapid decline in the number of clusters with age. However, these references use the term ``cluster'' in a wide sense, mostly referring to OB associations \citep[e.g.][]{pfalzner15}; see the discussion on terminology in \citet{krumholz14}.}
 for this apparent difference between the inner and outer regions of the Milky Way:
\begin{itemize}
\item The stronger gravitational potential in the inner Galaxy must be much more efficient at dissolving initially massive clusters over the course of 1~Ga.
\item High obscuration and crowding, prevalent towards the inner Galaxy, render many other existing massive cluster remnants essentially invisible to observers situated in our location. In support of this interpretation, the few populous moderately young open clusters known, such as M11 or NGC~6067 \citep{alonso17} are relatively nearby ($d\la2$~kpc) and in the middle of low-extinction windows. Despite this, they do not stand out strongly from their background. While younger populous clusters, such as Stephenson~2 \citep{negueruela12} or vdBH~222 \citep{marco14}, are easily identified as a collection of extremely bright infrared stars (their red supergiants), older clusters contain giants whose intrinsic brightness is comparable to the field red giant branch (RGB) or asymptotic giant branch (AGB) stars.
\end{itemize}

In this paper, we present evidence in the sense that many moderately young and intermediate-age clusters must remain unidentified because of background confusion by presenting a new, large cluster containing bright stars (about 30 members are brighter than $G=13$) that appears to be rather massive and has escaped detection until now. The new cluster, which we call Valparaiso~1, lies on the Galactic plane at $l=27\fd0$ and is thus projected on top of a line of sight that goes across the Sagittarius and Scutum-Crux arms and reaches the base of Scutum-Crux arm, close to the near end of the Bar \citep{negueruela12}. During a search for background red luminous stars \citep[reported in][]{negueruela12}, we found a large number of candidate early type stars in the vicinity of the catalogued emission-line star THA 14-29 (= TYC 5121-662-1) and the catalogued luminous star LS$\,$IV~$-05\degr$12. Some of these stars were observed spectroscopically, revealing a population of mid-B giants, highly indicative of a single-age population. \textit{Gaia} DR2 astrometric data confirm the existence of this population, while revealing a much larger extent and a high number of associated red giants. The cluster that we study is included in the list of cluster candidates presented by \citet{castrogin20}, under the name UBC~106. In this work, we use intermediate-resolution spectra of these red giants to determine their stellar parameters and \textit{Gaia} and 2MASS photometry to estimate cluster parameters.

\section{Observations and data}

Classification spectroscopy of blue stars in the area was obtained with the Boller \& Chivens Spectrograph on the 2.5-meter Ir\'en\'ee du Pont telescope at Las Campanas Observatory (Chile) during a run on 2012 June 25\,--\,27 June. The instrument was equipped with a Marconi CCD mated to a Bowen Schmidt camera as detector. We used the  blue 1200 l/mm grating, which gives a nominal dispersion of 0.8 \AA/pixel over a wavelength range of 163 nm. Together with a 2 arcsec slit, this grating results in a resolving power of $R=2\,450$.

Intermediate-resolution spectra of red giant candidates was taken on 2018 July 5\,--\,7 with the Intermediate Dispersion Spectrograph (IDS) mounted on the 2.5-m Isaac Newton Telescope (INT) at El Roque de los Muchachos
Observatory, in La Palma (Spain). We used the 235-mm camera with the \textit{Red+2} CCD and grating H1800V. This grating provides intermediate-high resolution over the whole optical range. In combination with a 1 arcsec slit, it gives an oversampled 2-pixel resolution element of 0.66\,\AA\ over an unvignetted range of $\sim640\,$\AA. We used it on July 5 and 6 centred on 6700\,\AA, achieving a resolving power, measured on arc lines, of $R\approx10\,000$. On July 7th, we centred the observed region on 8\,600\,\AA, obtaining $R\approx13\,000$ spectroscopy over the \ion{Ca}{ii} triplet range. 

Finally, on July 8th 2018, we used the H1800V grating centred on 4\,400\,\AA\ to take classification spectra of two further blue cluster members. By using a 1.4 arcsec slit, we obtained a resolving power of $R\approx6\,500$.

All the spectra have been reduced according to standard procedure with IRAF\footnote{IRAF is distributed by the National Optical Astronomy Observatories, which are operated by the Association of Universities for Research in Astronomy, Inc., under cooperative agreement with the National Science Foundation}. Bias removal, flat-fielding and wavelength calibration were done with frames taken on the same nights as the observations.

   \begin{figure}
   \centering
\resizebox{\columnwidth}{!}{\includegraphics[clip]{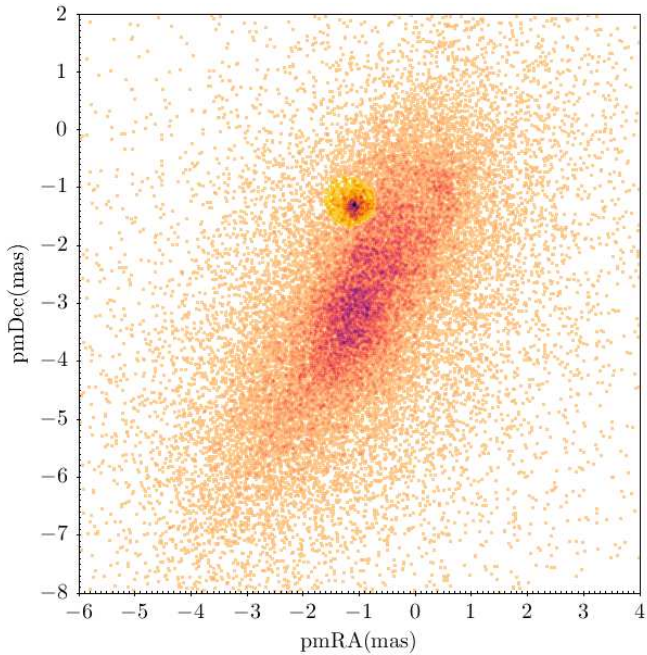}}
   \caption{Density map (linear scale) of the input catalogue for Clusterix in the proper motion plane. The yellow circle surrounding the densest clump, around ($-1.1$,$-1.3$) mas~$\mathrm{a}^{-1}$, indicates the region selected for the initial sample by the Clusterix analysis. \label{pmplane}}
    \end{figure}

\section{Results}

\subsection{Cluster definition}
\label{sec:def}

\textit{Gaia} DR2 data \citep{brown18} provide precise positions, proper motions and parallaxes for most of the stars brighter than $G=20$. Although there are systematic errors affecting all astrometric parameters \citep{luri18}, this is a high precision dataset that allows a very good definition of cluster membership. Since the cluster seems to be extended, we downloaded all the DR2 data within a circle of radius 50 arcmin around a nominal centre placed at RA: 18 41 52 Dec: $-05$ 27 00, in the midst of the blue stars detected. After removing objects with no astrometric solution or very large errors in the proper motions (error greater than $1\:$mas$\:\mathrm{a}^{-1}$ in any of the two ppms), we examined the proper motion plane, where the cluster is an obvious over-density around (pmRA $\equiv \mu_{\alpha} \cos \delta$, pmDec $\equiv \mu_{\delta}$) $\sim (-1.1,-1.3)\:$mas$\:\mathrm{a}^{-1}$ for any radius taken (cf. Fig.~\ref{pmplane}).

As this is a very crowded field with differential extinction, defining the extent of the cluster in the proper motion plane is not straightforward. Contamination by field stars is high, as can be seen in Fig.~\ref{pmplane}\footnote{The density of sources in the area is so high that the open cluster Teutsch~145, which is included in our circle, does not stand out at all in Fig.~\ref{pmplane}. This is a faint and distant cluster for which \citet{cantat18} identify around 100 likely members based on DR2 data and give (pmRA, pmDec)  $\sim (-1.64,-3.43)\:$mas$\:\mathrm{a}^{-1}$.}. To obtain an initial estimate, we used the Virtual Observatory tool Clusterix 2.0 \citep{balaguer20}. Clusterix is an interactive web-based application to calculate the grouping probability of a list of objects by using proper motions and the non-parametric method proposed by \citet{cabreracano90} and described in \citet{galadi98}. In its current version, Clusterix works only in the proper motion plane, ignoring all other information. We run the tool interactively by defining concentric circles of changing radius around the nominal centre. For ease of computation, we restricted ourselves to DR2 objects with $G\leq16$ and errors in proper motion below $1\:$mas$\:\mathrm{a}^{-1}$. The best results were obtained for a cluster of radius 9 arcmin and a field located between 36 arcmin from the centre and the edge of the dataset. Based on an empirical determination of the frequency functions in the vector point diagram \citep{sanders71}, Clusterix assigns each object a probability of belonging to a distinct population, identified as the cluster.

   \begin{figure*}
   \centering
\resizebox{\columnwidth}{!}{\includegraphics[clip]{sel_position.eps}}
\resizebox{\columnwidth}{!}{\includegraphics[clip]{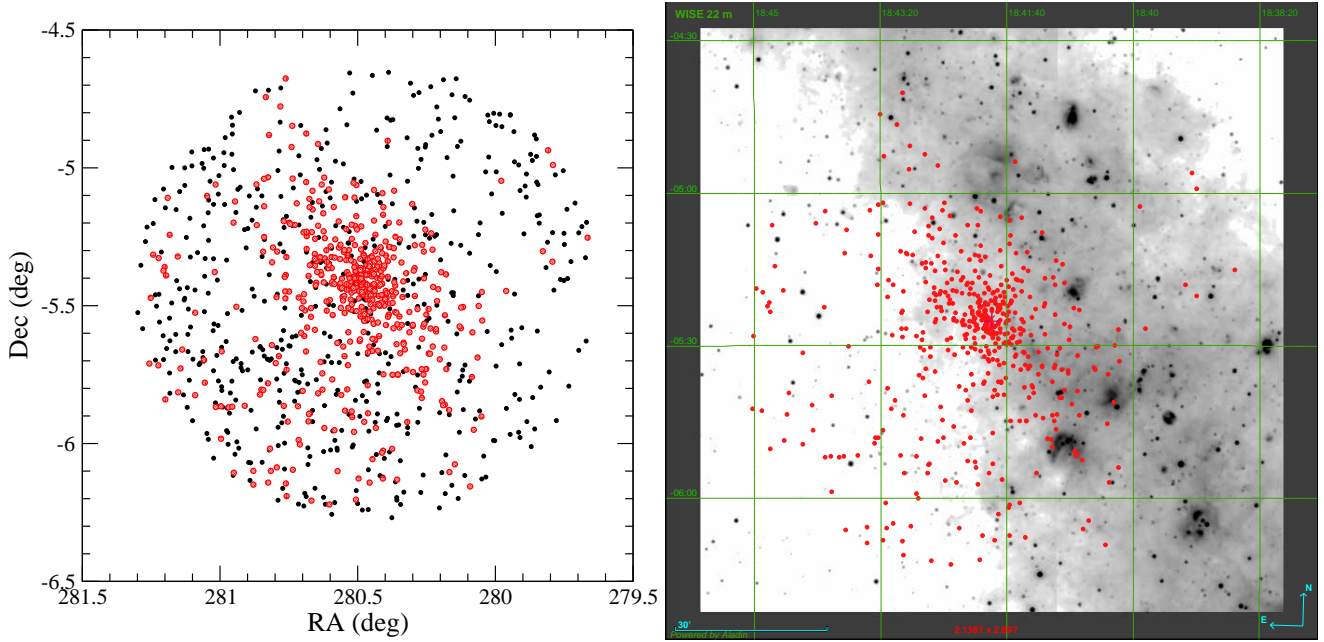}}
   \caption{\textit{Left panel: }Spatial distribution of candidate cluster members selected from \textit{Gaia} data. The black dots represent the initial selection from the probabilities assigned by Clusterix based on proper motions only. The small red circles are the result of the cleaning procedure described in the text, by using parallaxes and \textit{Gaia} photometry (i.e. position in the CMD).  \textit{Right panel: } The same clean sample is shown on top of the WISE 22~$\mu$m map of the region. The anticorrelation between likely member density and dust emission is strong. \label{spatial}}
    \end{figure*}
%
%

In fact, Clusterix identifies two distinct populations within the area. The first one, representing a weak overdensity around (pmRA, pmDec) $\sim (+0.5,-1.0)\:$mas$\:\mathrm{a}^{-1}$ is not strongly concentrated towards the cluster position, but rather extends over the whole field, with a higher frequency around RA: 18:41:15, Dec: $-$05:26:20, a region that seems to show a gap in obscuration. Our data do not allow the identification of this population as a cluster. Nevertheless, \citet{castrogin20}, by using automated non-parametric techniques, have recently identified a high-confidence cluster candidate in this area (UBC~105) with astrometric parameters pmRA = $0.46\pm0.11\:$mas$\:\mathrm{a}^{-1}$, pmDec = $-0.99\pm0.09\:$mas$\:\mathrm{a}^{-1}$, which seems to correspond exactly to this population. Therefore we assume that this is a real cluster for which we will use the designation UBC~105 throughout.

\begin{figure}
   \centering
\resizebox{\columnwidth}{!}{\includegraphics[,clip]{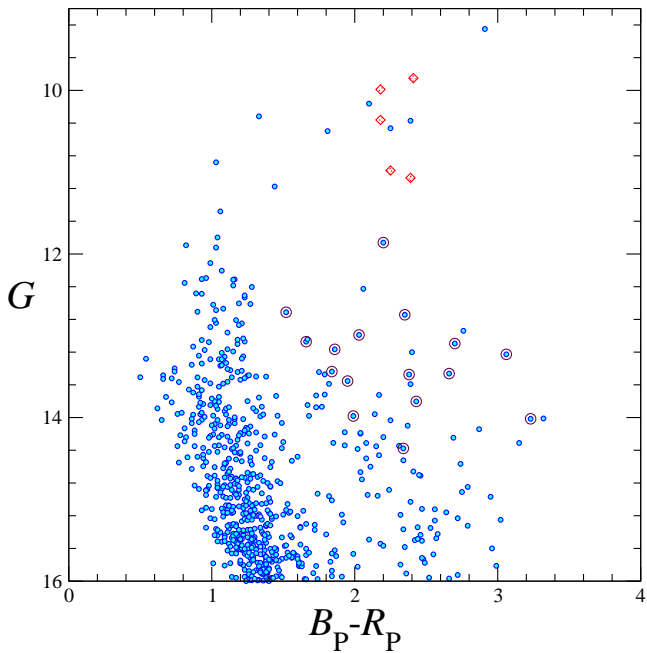}}
   \caption{\textit{Gaia} CMD for the sample selected using astrometric data. The diamonds indicate the five bright stars with compatible RVs. Large circles show other stars with \textit{Gaia} RVs within the sample. Only one of them has a RV compatible with the brighter group.\label{gaiaphot}}
    \end{figure}

The second, much better defined, overdensity -- easily detected by eye in Fig.~\ref{pmplane} -- is strongly concentrated towards the area where we had previously identified a group of early-type stars. After analysing the correlation between probability assigned by Clusterix and distribution in the proper motion plane, we made a cut in probability (which is given in arbitrary units) that separates the region shown in Fig.~\ref{pmplane}. The population selected consists of 1047 objects (out of a total population of over 35\,500 objects). The spatial distribution of this selection can be seen in Fig.~\ref{spatial} (left panel), where the objects are marked as black dots, and very clearly demonstrates its clustered nature. However, given the high density of field stars over the whole extent of Fig.~\ref{pmplane}, the sample must still contain substantial contamination. We performed an initial cleaning by making use of \textit{Gaia} parallaxes. The sample has a mean and median parallax of $\pi = 0.39$~mas and a mode $\pi = 0.40$~mas, with standard deviations of $0.05$~mas. This excellent agreement confirms the existence of a well-defined population. We analysed the distribution of objects in the pmRA/plx and pmDE/plx plane to identify the locus of the population, and then proceeded to clean iteratively the sample of outliers. Next, we plotted the \textit{Gaia} CMD for the remaining sample (678 stars), which is shown in Fig.~\ref{gaiaphot}. The CMD confirms that most of the stars selected are part of a single population, with a typical stellar sequence around $BP-RP\approx1.2$, although considerably broadened by differential reddening, and an apparent clump of bright, red objects. The reality of this clump of evolved stars is substantiated by \textit{Gaia} radial velocities \citep[RVs;][]{katz19}. There are five stars brighter than $G\approx11$ that have \textit{Gaia} RVs, and they are all consistent with a given value (see Table~\ref{partab}). Nevertheless, substantial contamination can be seen at fainter magnitudes for $BP-RP>1.6$. In addition, there is a hint of a second population around $G\approx13.5$ with $BP-RP<1.0$. Closer examination, however, shows that all these objects with bluer colours are located in the Southeast quadrant, where extinction seems to be noticeably lower. They are then in all likelihood less reddened cluster members.

In view of this, we clean our sample by removing all stars with $G>11.5$ and $BP-RP>1.5$. 
The median astrometric parameters of the remaining sample are (pmRA, pmDec) = ($-1.08\pm0.11$,$-1.32\pm0.11$)\:mas$\:\mathrm{a}^{-1}$, $\pi=0.40\pm0.05$~mas, where the errors represent the standard deviations of the sample. After removal of a handful of outliers, we are left with a sample of 479 objects down to $G=16$, compatible within their errors at $2\,\sigma$ with these average values. Their spatial distribution, also showed in Fig.~\ref{spatial}, is much more concentrated than that of the initial sample. The almost complete absence of objects in the Western third of the circle correlates strongly with the presence of dark clouds in the DSS images of the area (cf. Fig.~\ref{spatial}, right panel). 

%
              %
%

We used the Automated Stellar Cluster Analysis (ASteCA) code \citep{perren} to explore the cluster's geometry.  Running ASteCA on both the initial and cleaned samples displayed in Fig.~\ref{spatial} (left panel), we fit a King profile. The centre of the cluster is found at  RA = 280\fdg47 (18 41 53),  Dec= $-5\fdg$42 ($-05$ 25 12), about 2 arcmin to the North of our initial guess. The radius is given as $r=0.15\pm0.02$~deg, but this is clearly an underestimation, as the "background" in these samples is made of the same population. 

For an independent confirmation, we took 2MASS data within 20 arcmin of this centre. We retained only stars with ``good'' quality flags (A or E) in all three filters and removed all stars with high errors by taking 0.1~mag in the $(J-K_{\mathrm{S}})$ colour as a limit. We selected early type stars by making use of the infrared $Q$ parameter \citep[see][]{ns07}. We considered a range $-0.14 < Q_{\mathrm{IR}} < 0.08$ to include the region inhabited by early-type (OBA) stars and take into account the typical photometric errors. We rejected stars brighter than $K=9$, as these will almost certainly be foreground objects. This procedure leaves 1864 candidate early-type stars. Again ASteCA detects a strong overdensity, this time centred on  RA = 280\fdg49, Dec = $-5\fdg$41, about one arcmin to the northeast of the centre determined from \textit{Gaia} candidates. Given that the sample and method are completely independent from the \textit{Gaia} data, this confirms the definition of the cluster. The King profile, however, does not result in a good fit, as the spatial distribution is very asymmetric. Again, the reason for this is evident in Fig.~\ref{spatial} (right panel).  There is a number of dark regions immediately to the West of the cluster core that significantly reduce stars counts to that side. In particular, the density maps of both \textit{Gaia} sources and 2MASS sources show a strip starting close to the cluster centre and extending to the southwest for about half a degree with a much lower source count than neighbouring areas. Optical images show that this strip corresponds to a dark lane, which must be foreground to the cluster. In view of these complications, we do not attempt to define a real cluster centre, although we note that members are very strongly concentrated towards the position used as nominal centre.

Finally, we note that \citet{castrogin20} identify a high-confidence cluster candidate in this area, UBC~106, with astrometric parameters pmRA = $-1.10\pm0.10\:$mas$\:\mathrm{a}^{-1}$, pmDec = $1.32\pm0.11\:$mas$\:\mathrm{a}^{-1}$, $\pi$ = $0.40\pm0.04\:$mas. Given the total correspondence in all parameters, this cluster candidate UBC~106 is identical to our Valparaiso~1, offering independent confirmation of its existence and characteristics.

%

%
              %
%

\subsection{Stellar content}

\begin{figure*}
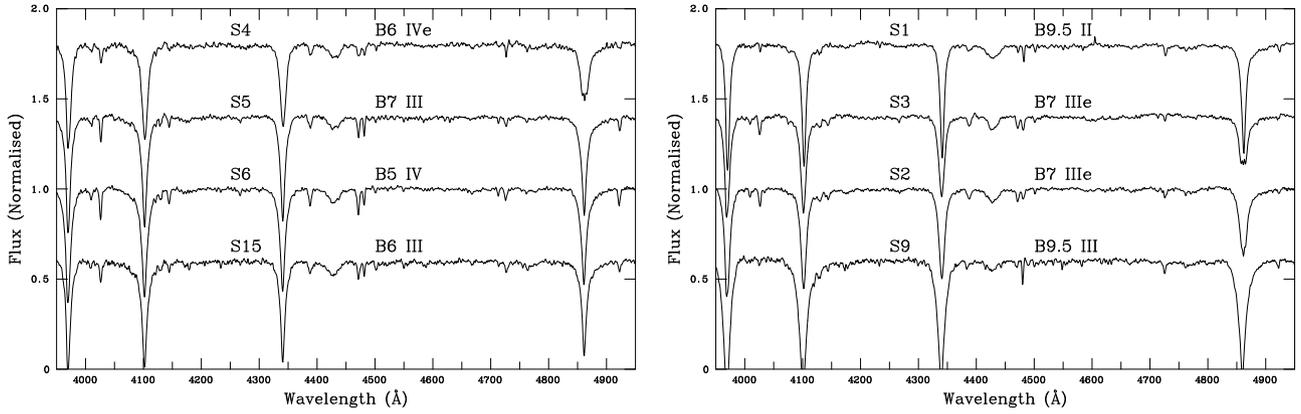

   \centering
\resizebox{\columnwidth}{!}{\includegraphics[angle=-90,clip]{core_blue.eps}}
\resizebox{\columnwidth}{!}{\includegraphics[angle=-90,clip]{others_blue.eps}}
   \caption{Intermediate-resolution spectra of blue stars in the field of Valparaiso~1. \label{blues}}
    \end{figure*}

Classification spectra of stars observed with the DuPont telescope are shown in Fig.~\ref{blues}. Classification was performed using classical criteria by comparison to a set of spectra of standard stars (as discussed in \citealt{negueruela19}) degraded to the same resolution. Since the target selection was made at the time by using only 2MASS CMDs, there were a few F-type interlopers (not shown). The spectral classification of the B-type stars is shown in Table~\ref{classtab}, together with their \textit{Gaia} DR2 astrometric parameters.  The majority of the stars selected turn out to be B5\,--\,7 giants and subgiants, as is typical of a moderately young cluster. LS~IV $-05\degr$12  is brighter, and at B9.5\,II falls a bit short of supergiant luminosity. TYC 5121-570-1 has a spectral type B2\,IV and different proper motions, so we consider it a clear non-member. Star~9 has proper motions compatible with membership, but its parallax ($\pi = 0.68$) and spectral type B9.5\,III suggest it is a foreground object.  All the other stars have very similar spectra and hence spectral classifications. TYC 5121-533-1, however, is far from the central concentration and has very different proper motions. In fact, according to \citet{castrogin20}, TYC 5121-533-1 is the second brightest likely member of UBC~105.

All the other stars have astrometric parameters compatible with cluster average to within three sigma, and so we do not discard their membership. TYC~5121-381-1 has a high parallax ($\pi=0.65$), but this is a Be star, and so this value could be affected by the presence of an envelope. In fact, its EDR3 parallax \citep{lindegren20} is a much shorter $0.56\pm0.03$. We keep it as a possible member until further \textit{Gaia} releases. Two other stars in the sample are also clearly Be stars. We note that Be stars with late-B types tend to have weak emission features \citep[cf.][]{steele99} and so we could be missing some other Be stars that are only detectable in H$\alpha$ (as we have not observed this spectral range). The brightest star in our sample, LS  IV $-05\degr$12, is not picked as a member by \citet{castrogin20} because of its poor DR2 astrometric parameters, but its EDR3 values bring it close to the cluster average. 

              %
%
%
To check the validity of the astrometric selection adopted to define the cluster, we observed two of the brightest blue stars chosen as astrometric members outside the central concentration with the INT + IDS. These two stars (101 and 109 in Table~\ref{classtab}) have the same spectral types as the members of similar brightness in the central concentration, confirming that the cluster extends well beyond the central 5 arcmin. Star 109 may be a mild Be star, but an  H$\alpha$  spectrum will be needed to confirm this.

\begin{table*}
        \centering
        \caption{Observed parameters for blue stars with classification spectra.}
        \label{classtab}
        \begin{tabular}{lcccccccccc}
                \hline
                \noalign{\smallskip}
\textit{Gaia} &RA &  Dec & Other & Spectral & pm (RA) & pm (Dec)& $\pi$  & $G$ & $BP$-$RP$ & Member$^{\dagger}$\\ 
&&& name & type & (mas) & (mas) & (mas) & (mag) & (mag) \\
\noalign{\smallskip}
\hline
\hline
\noalign{\smallskip}%
1   &18:41:52.26 &   $-$05:23:22.1 &LS  IV $-05\degr$12 & B9.5\,II  & $-0.75\pm0.14$ &   $-1.12\pm0.13$ &  $0.43\pm0.07$& 10.32& 1.32&L   \\
2   &18:41:52.13 &   $-$05:26:58.1 &TYC 5121-381-1 & B7\,IIIe&        $-1.06\pm0.09$ &   $-1.53\pm0.09$ &  $0.64\pm0.05$& 11.92& 1.11&P   \\
3   &18:41:47.13 &   $-$05:27:55.2 & TYC 5121-662-1  & B7\,IIIe &     $-0.86\pm0.08$ &   $-1.66\pm0.08$ &  $0.35\pm0.05$& 11.85& 0.98&L   \\
4   &18:41:55.90 &   $-$05:26:58.5 & & B6\,IVe &		      $-1.10\pm0.08$ &   $-1.30\pm0.08$ &  $0.41\pm0.05$& 12.53& 1.23&C   \\
5   &18:41:52.27 &   $-$05:26:08.6 & & B7\,III &		      $-1.16\pm0.07$ &   $-1.35\pm0.07$ &  $0.37\pm0.04$& 12.60& 1.19&C   \\
6   &18:41:52.50 &   $-$05:25:15.1 && B5\,IV &  		      $-0.97\pm0.07$ &   $-1.39\pm0.07$ &  $0.44\pm0.04$& 12.51& 1.23&C   \\
8   &18:41:48.20 &   $-$05:26:46.2 && B5\,IV &  		      $-0.42\pm0.08$ &   $-0.53\pm0.07$ &  $0.48\pm0.05$& 12.42& 0.90&P   \\
9   &18:41:47.66 &   $-$05:25:53.1 & & B9.5\,III&		      $-0.82\pm0.08$ &   $-1.28\pm0.08$ &  $0.68\pm0.05$& 12.23& 1.18&U   \\
11  &18:41:32.80 &   $-$05:25:15.5 &TYC 5121-570-1 & B2\,IV&	      $-0.91\pm0.09$ &   $-3.69\pm0.08$ &  $0.45\pm0.05$& 11.70& 0.86&N   \\
13  &18:41:26.06 &   $-$05:24:54.9 &TYC 5121-533-1& B5\,III&	      $+0.42\pm0.10$ &   $-0.97\pm0.09$ &  $0.49\pm0.05$& 11.97& 1.00&N   \\
15  &18:41:20.28 &   $-$05:24:43.1 &TYC 5121-63-1& B6\,III&	      $-1.04\pm0.07$ &   $-1.28\pm0.07$ &  $0.42\pm0.04$& 11.92& 1.03&C   \\
101 &18:41:27.37 & $-$05:31:23.1& & B7\,III &  $-1.22\pm0.07$ & $-1.31\pm0.07$ & $0.43\pm0.04$ & 11.80 & 1.04 & C\\
109 &18:41:51.14 & $-$05:19:55.6 & & B6\,IIIe? & $-1.13\pm0.09$ & $-1.27\pm0.08$ & $0.40\pm0.05$ & 12.31 & 1.15 & C\\

\noalign{\smallskip}
                \hline
        \end{tabular}
	\begin{list}{}{}
\item[]$^{\dagger}$ Membership is considered C(ertain) for stars with astrometric parameters compatible with the cluster average and appropriate spectral types; (L)ikely for stars with appropriate spectral types whose astrometric parameters are compatible at $2\sigma$ within their respective errors; (P)ossible for a star with appropriate spectral type and the right parallax, but divergent proper motion; (U)nlikely for a star with divergent parallax and a spectral type B9.5\,III, which does not agree with the rest of members. The B2\,IV star and the member of UBC~105, Star 13, are considered certain (N)on-members.
\end{list}
\end{table*}

\begin{figure}
   \centering
\resizebox{\columnwidth}{!}{\includegraphics[angle=0,clip]{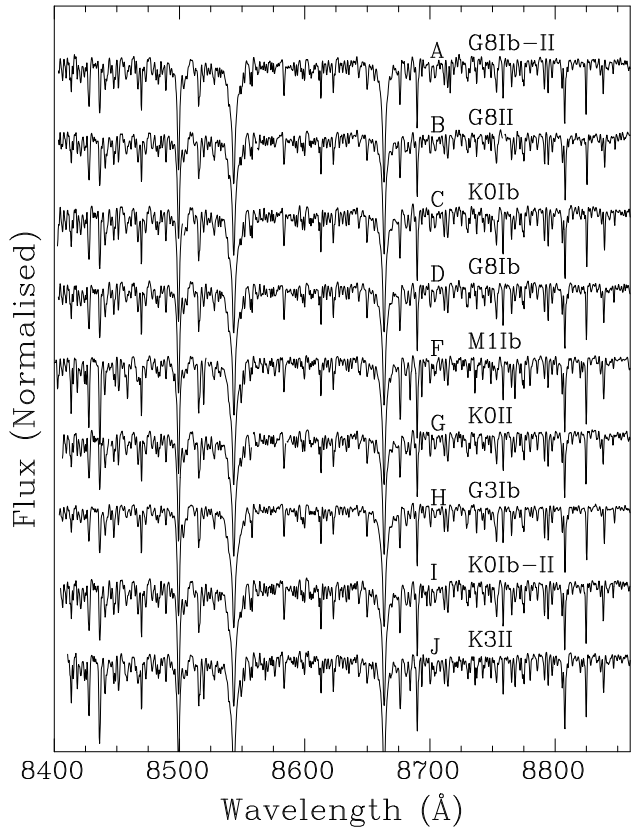}}
   \caption{Spectra of nine cool luminous members of Valparaiso~1, used for spectral classification. \label{rgs}}
    \end{figure}

The nine objects occupying positions compatible with being luminous cool stars in Fig.~\ref{gaiaphot} were observed in the \ion{Ca}{ii} triplet (CaT) range for classification purposes. Criteria to classify late-type stars in this range are discussed in \citet{negueruela11} and \citet{dorda16}. Their spectra, displayed in Fig.~\ref{rgs}, confirm that they are all luminous cool stars, straddling the limit between bright giants and supergiants, as seen in moderately young (i.e. $\sim50$\,--\,$100\:$Ma) clusters \citep[e.g.][]{negmar12, alonso17, negueruela18}.

\subsection{Spectroscopic analysis}

The spectra of cool luminous stars can be used to determine stellar properties. For the ten objects listed in Table~\ref{partab} we carried out such an analysis.
Because of the poor weather during the night and low response of the CCD in the \ion{Ca}{ii} region, these triplet spectra have a low signal-to-noise ratio and we used them for spectral classification only. We used the spectra in the H$\alpha$ region to compute effective temperature ($T_{\textrm{eff}}$), surface gravity ($\log{g}$), and metallicity ([M/H]) by comparing them to a grid of synthetic spectra \citep[as in ][]{lohr18}. We employed the new version of the code {\scshape SteParSyn} \citep[see][ Tabernero et al., in prep.]{tabernero18,tabernero21}, which uses a Markov-chain Monte Carlo algorithm \citep[emcee, see][]{emcee} for optimisation. We explored the parameter space by using 12 Markov-Chains of 1\,000 points. 

We generated a grid of synthetic spectra by means of the MARCS spherical atmospheric models \citep{gus08} and the radiative transfer code \textit{spectrum} \citep{graco94}. As line list, we employed a selection from the VALD3 database \citep{pis95,kup00,rya15}, which takes into account all the relevant atomic and molecular features (dominated by TiO and CN) that can appear in cool luminous stars earlier than mid-M types. We employed the Anstee, Barklem, and O'Mara theory (ABO) as Van der Waals damping prescription, when available in VALD3 \citep[see][]{bar00}. The grid spans $T_{\rm eff}$ from $3\,500\:$K to $8\,000\:$K with a step of $250\:$K above $4\,000\:$K and $100\:$K otherwise. The metallicity ranges from [M/H]$=-1.0\:$dex to [M/H]$=1.0\:$dex in $0.25\:$dex steps, while surface gravity goes from $-0.5$ to 2.0~dex in 0.5~dex steps, when available in the MARCS grid. The microturbulence ($\xi$) was adjusted according to the 3D model based calibration described in \citet{dut16}. We convolved our grid of synthetic spectra with a Gaussian kernel to account for the line-spread function  of the instrument. The results of the analysis are listed in Table~\ref{partab}.


\begin{table*}
        \centering
        \caption{Stellar parameters for cool stars in the field of Valparaiso~1.}
        \label{partab}
        \begin{tabular}{lccccccccc}
                \hline
                \noalign{\smallskip}
Star &  Other & \textit{Gaia} & Spectral & $T_{\textrm{eff}}$ & $\log\,g$ &  [M/H] & $v_{{\rm hel}}$ & RV (\textit{Gaia})\\ 
& name & DR2 & type & (K) & & (dex) & (km~s$^{-1}$) & (km~s$^{-1}$) \\
\noalign{\smallskip}
\hline
\hline
\noalign{\smallskip}%
A & TYC 5121-543-1 & 4256511915482900608 & G8\,Ib--II &$4693 \pm    46$ & $1.1 \pm  0.11$ & $-0.05 \pm  0.06$ &  $40.6\pm0.2$& $-$\\ 
B & GSC 05121-00622& 4256512843232515840 & G8\,II & $4620 \pm    51$ & $ 1.42 \pm  0.12$ & $ +0.01 \pm  0.07$ & $42.4\pm0.2$ & $40.1\pm0.4$\\
C & TYC 5121-819-1 & 4253508943153458048 & K0\,Ib & $4639 \pm    40$ &  $0.88 \pm  0.11$ & $+0.02 \pm  0.06$ & $39.8\pm0.2$ & $-$\\
D &  TYC 5121-218-1 & 4253508702635208832 & G8\,Ib &$4640 \pm    48$ & $ 1.02 \pm  0.11$ & $ -0.07 \pm  0.06$ &  $41.1\pm0.2$& $40.8\pm0.5$\\
E &  CM Sct & 4253603501158148736 & $-^{\dagger}$ & $5431 \pm    36$ & $ 1.03 \pm  0.09$ &  $-0.15 \pm  0.04$  & $47.4\pm  0.3$& $-$\\
F &  TYC 5121-758-1 & 4253603501158148736 & M1\,Ib & $3840 \pm    20$  & $0.33 \pm  0.09$ & $-0.10 \pm  0.05$ &  $41.6\pm0.2$& $-$\\
G & & 4256511468842481408 & K0\,II& $4725 \pm    44$ & $ 1.33 \pm  0.09$ & $+0.12 \pm  0.05$ &  $41.5\pm0.2$ & $41.6\pm0.4$\\
H & TYC 5121-684-1 & 4253603501158148736 & G3\,Ib & $5105 \pm    27$ &  $0.72 \pm  0.08$ & $-0.07 \pm  0.04$ &  $41.1\pm0.2$ & $41.8\pm0.2$\\
I & TYC 5125-1531-1 & 4253499219346450432 & K0\,Ib-II& $4755\pm22$ & $0.91 \pm 0.06$ & $+0.08 \pm 0.03$ & $43.5\pm0.6$ & $42.1\pm0.3$\\
J & & 4253597556923196672 & K3\,II & $4137 \pm    40$ &  $0.65 \pm  0.1$ & $-0.06 \pm  0.06$ &  $43.4\pm0.2$ & $-$\\
\noalign{\smallskip}
                \hline
        \end{tabular}
	\begin{list}{}{}
\item[]$^{\dagger}$ CM Sct is a classical Cepheid and thus spectral variable. Therefore we did not observe it in the CaT range.
\end{list}
\end{table*}

The metallicities derived for all the cool stars are roughly compatible with a solar metallicity. The only exception is the Cepheid variable CM~Sct, which gives a decidedly subsolar metallicity. This is in open contrast to the slightly supersolar values reported by \citet{luck18}. A similarly large discrepancy is seen between this authors' value for $\log g$ (2.1) and ours (1.0). Since metallicity and effective gravity are highly degenerate, we excluded CM~Sct from the calculation of the average metallicity, which, after weighing each value with its error, is exactly solar. In view of this, we use solar-metallicity isochrones in the subsequent analysis.

RVs were measured by cross-correlating each individual observation against the Arcturus atlas \citep{hinkle00} by means of the {\sc iSpec} code \citep{blancua14}. Comparison of our RVs to \textit{Gaia} DR2 values for five stars in common shows excellent agreement. Our formal errors are clearly an underestimation, as systematic effects preclude errors better than $\sim3\:$ km~s$^{-1}$ for an instrument with such large flexures as IDS. All ten stars display consistent values of RV. Only CM~Sct appears slightly deviant, but this well-studied Cepheid is known to display RVs ranging from $\sim+23$ to $\sim+55\:$km~s$^{-1}$ \citep{metzger91}, with a systemic velocity of $+40.8\:$km~s$^{-1}$ \citep{pont94}, fully consistent with the others. Removing the variable, our average RV value is $+41.7\:$km~s$^{-1}$ ($\sigma = 1.2\:$km~s$^{-1}$ for nine stars), while the \textit{Gaia} DR2 unweighted average is  $+41.3\:$km~s$^{-1}$ ($\sigma = 0.8\:$km~s$^{-1}$ for five stars).

\section{Cluster parameters}

\begin{figure}
   \centering
\resizebox{\columnwidth}{!}{\includegraphics[angle=0,clip]{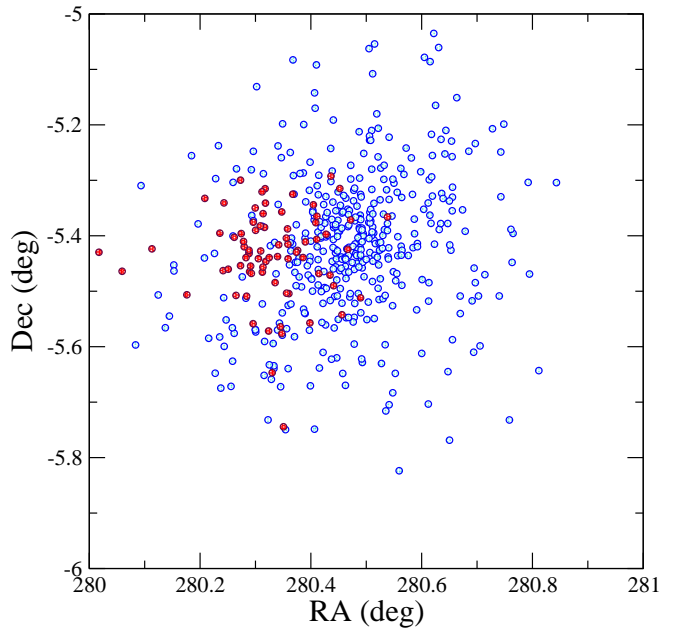}}
   \caption{Spatial distribution of likely members (according to \citealt{castrogin20}) of UBC 106 = Valparaiso~1 (blue circles) and UBC~105 (dashed red circles). Although the two clusters overlap, Valparaiso~1 is much more populous and extends over a vast area that encompases UBC~105.   \label{blended}}
    \end{figure}

The derivation of cluster parameters is complicated. The cluster is large on the sky and extinction varies strongly across its surface. The photometric diagrams are complicated by the presence of other populations. In particular, UBC~105 is another young cluster at about the same distance ($\pi=0.47\pm0.03\:$mas; \citealt{castrogin20}). Although the two clusters are very well separated in the proper motion plane, they are spatially coincident, as seen in Fig.~\ref{blended}.

\begin{figure}
   \centering
\resizebox{\columnwidth}{!}{\includegraphics[angle=0,clip]{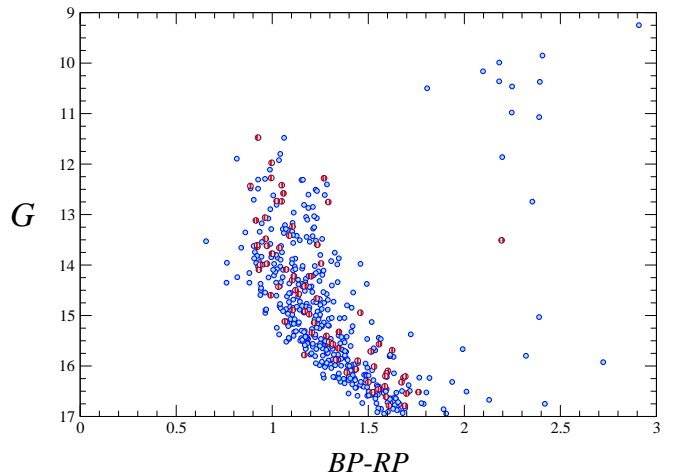}}
   \caption{Colour magnitude diagram, using \textit{Gaia} DR2 photometry of likely members \citep{castrogin20} of Valparaiso~1 (blue circles) and UBC~105 (dashed red circles). The two populations occupy indistinguishable positions in this diagram. \label{cmdblend}}
    \end{figure}

    Moreover, the two clusters have similar reddening, and their populations blend in all photometric diagrams. As seen in Fig.~\ref{cmdblend}, they must have about the same age, as also suggested by the fact that TYC~5121-533-1, one of the brightest members of UBC~105, has the same spectral type as the brightest members of Vaparaiso~1.

\begin{figure}
   \centering
\resizebox{\columnwidth}{!}{\includegraphics[angle=0,clip]{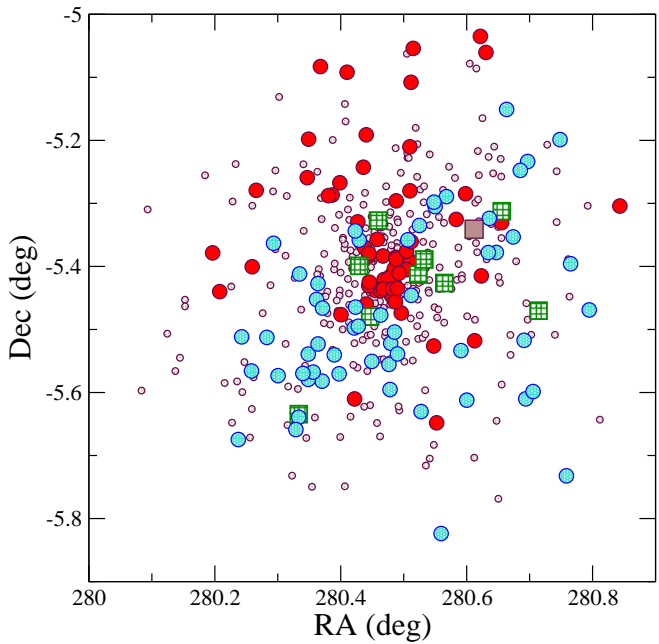}}
   \caption{Spatial distribution of highly-reddened (red circles) and less reddened (blue circles) early-type members of Valparaiso~1. Fainter members are shown as smaller purple circles. Cool luminous stars appear as green squares, while the Cepheid CM~Sct is the filled brown square.  \label{twobranches}}
    \end{figure}

The effect of differential extinction is obvious in Fig.~\ref{cmdblend}. Astrometric cluster members of a given brightness have ($BP-RP$) colours ranging over more than half a magnitude. In fact, it is possible to identify two almost separate branches in the upper part of the CMD. To investigate their meaning, we divided blue members of Valparaiso~1 brighter than $G\approx14.5$ into two groups, according to their position in the CMD. Stars with  $BP-RP> 1.05$ were assigned to the high-reddening group, while stars with lower values were tagged as low reddening. Fig.~\ref{twobranches} shows the spatial distribution of the two groups. Although there is a fair amount of blending, the spatial segregation is strong. All the stars in the core region belong to the high-reddening group, together with the population to the northwest. This distribution clearly shows that the clouds visible in the optical images contribute large amounts of obscuration to the cluster core.

    \subsection{\textit{Gaia} photometry}
    
    In view of these difficulties, we have decided to carry out the analysis by utilising only the list of likely members determined by \citet{castrogin20}, as their criteria are stricter than ours -- we identify 479 objects down to $G=16$ as high-probability members, while they only list 431 members down to $G = 17$.  Of the B-type stars listed in Table~\ref{classtab}, only those that we chose as certain members are included in their list. None of our likely or possible members is. As seen in Fig.~\ref{cmdblend}, less than ten of the 431 astrometric members occupy positions incompatible with membership in the \textit{Gaia} CMD. However, the width of the main sequence noticeably decreases for $G>15$. Given that reddening is expected to affect equally stars of all magnitudes, this effect is very likely suggesting that many faint stars are being rejected as high probability members by the procedure employed by \citet{castrogin20} because of large astrometric errors (note that this effect is not present in the diagram presented in Fig.~\ref{gaiaphot}, where we select likely members simply by taking astrometric parameters within 2\,$\sigma$ of the cluster averages). In consequence, we cannot consider the list of members that we are using complete, at least for faint magnitudes. The reddening to the evolved members is more difficult to measure, but their spatial distribution (shown in Fig.~\ref{twobranches}) suggests that we can expect to find them at both high and low reddening.

\begin{figure}
\resizebox{\columnwidth}{!}{\includegraphics[angle=0,clip]{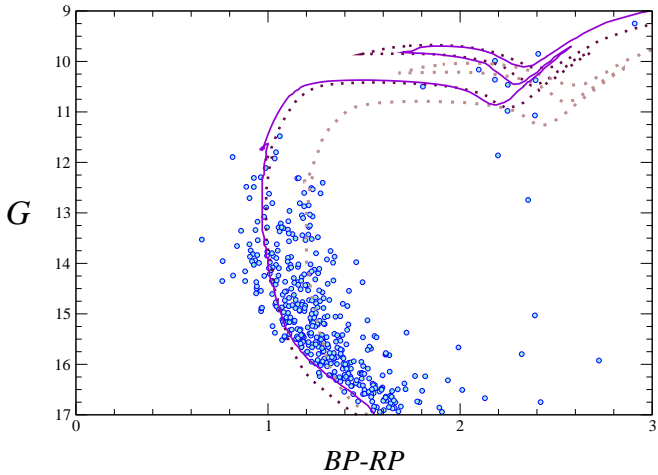}}
   \caption{Possible {\sc parsec} isochrone fits for the \textit{Gaia} DR2 CMD of Valparaiso~1. The solid line is a 90~Ma isochrone reddened with $A_V = 2.6$ and displaced to $DM = 11.6$. The dashed line is a 70~Ma isochrone with $A_V=2.7$ displaced to $DM=12.0$. Finally, the dotted line is the same 70~Ma isochrone with $A_V = 3.2$ displaced to $DM=11.8$. No 90~Ma isochrone can fit the high-reddening branch for a distance compatible with the \textit{Gaia} parallax. \label{cmd_isos}}
    \end{figure}

    With such strong differential reddening, isochrone fitting becomes very difficult. We can use the \textit{Gaia} DR2 parallax to provide an estimate of the distance, although being aware of the possibility of large systematic errors \citep{luri18}. A $\pi=0.40$, together with a global systematic error of $-0.03$, implies a distance of 2.3~kpc, i.e., a distance modulus $DM = 11.8$. In Fig.~\ref{cmd_isos}, we show a number of isochrones that can fit part of the cluster population. We downloaded isochrones from the Padova {\sc parsec} server \citep{parsec}, selecting a Kroupa IMF corrected for binaries and the \textit{Gaia} passbands from \citet{mapw18}. A 70~Ma isochrone reddened with $A_V=2.7$ can fit the upper main sequence and the cool luminous giants for $DM=12.0$. However, it fails to fit the fainter part of the main sequence. Given that there seems to be a substantial lack of members in this region, the importance of this failure is difficult to assess. A 90~Ma isochrone reddened with $A_V = 2.6$ gives a very good fit to the less reddened main sequence, and provides a decent fit to the evolved stars. Much higher extinction is needed to fit the more heavily reddened population. The 90 Ma isochrone requires $A_{V}=3.1$ to fit the more reddened blue members. But then a $DM$ no longer than $11.2$~mag is required to reach the position of the evolved stars. This would be equivalent to $\pi= 0.59$~mas, a value incompatible with the \textit{Gaia} DR2 determination, even if all sources of uncertainty are taken into account. On the other hand, the 70~Ma isochrone requires $A_V = 3.2$ to fit the stars from the core region, and reaches the position of the evolved stars for $DM$ in the 11.6 to $11.8$~mag range, as illustrated in Fig.~\ref{cmd_isos}. Therefore, if we assume a single age and give value to the \textit{Gaia} DR2 distance, the older age is strongly disfavoured and we are left with an age in the range 70 to 80~Ma, and a distance modulus in the 11.6\,--\,12.0~mag range ($ 2.1\:\mathrm{kpc}\, \la d \la\, 2.5\:\mathrm{kpc}$).
    
    Along this line of sight, the Sagittarius arm is believed to run almost perpendicularly at around 2~kpc \citep[see][for a detailed discussion]{marco11}. Therefore the shorter distance is favoured in terms of Galactic structure. We note the presence of the similarly aged open clusters NGC~6664 and NGC~6649 at about the same distance \citep{alonso20} in the same region of the sky. The older clusters M11 and NGC~6704, which are very close in the sky to Valparaiso~1, are also located at about the same distance \citep{cantat18}. Interestingly, within 2 degrees of Valparaiso~1, \citet{castrogin20} find five cluster candidates with very similar parallaxes. Apart from UBC~105, discussed above, UBC~109 and UBC~349 are also well populated clusters, whose \textit{Gaia} DR2 CMD suggest somewhat older ages than Valparaiso~1, while UBC~353 and UBC~355 are poorly populated candidates.
    
    Finally, we note that \citet{cantat20}, by using a machine-learning automated procedure, derive a distance modulus $DM=11.9$ and age of 160~Ma for UBC~106, which we identify with Valparaiso~1. However, this fit is obtained for an extinction $A_{V}$ of only $1.9$~mag, suggesting that the automated procedure has only taken into account the small subset of less reddened members of the cluster.
    
   \subsection{2MASS photometry}
   
    For a complementary analysis, we used near-infrared photometry. We obtained $JHK_{{\mathrm S}}$ photometry from the 2MASS catalogue \citep{skru06}. The completeness limit of this catalogue is set at $K_{{\mathrm S}}=14.2$. Nevertheless, in this crowded region of the Galactic Plane, a very large fraction of objects are affected by crowding, resulting in bad quality flags and measurements marked as upper limits (flag U) even for bright objects. In consequence, we did not attempt to use the whole 2MASS dataset for the region, but rather only the photometry available for the objects selected from \textit{Gaia} DR2 data. We performed a cross-match over the whole region of interest between the members selected by \citet{castrogin20} and 2MASS, using a match distance of only 0.2 arcsec, which seems appropriate given the high density of 2MASS sources and the accuracy of astrometry in both catalogues. Even so, the resulting selection contains a very large number of faint objects close to the detection limit of 2MASS. In consequence, the list was cleaned by removing all objects with poor photometry (upper limits or errors higher than 0.1~mag) in any of the three filters. Although almost two thirds of the sources are rejected, we are still left with 150 objects with good photometry. 

\begin{figure}
\resizebox{\columnwidth}{!}{\includegraphics[angle=0,clip]{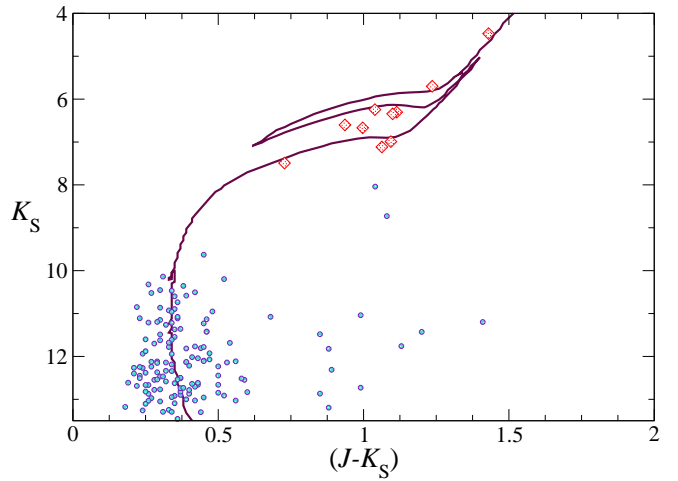}}
   \caption{Near-IR CMD for Valparaiso~1. The blue circles represent 2MASS cross-matches for members according to \citet{castrogin20} that have well-defined photometric errors. The diamonds indicate the ten astrometric and RV cool luminous members. A representative isochrone is shown, with the preferred parameters from the optical-data fit (75~Ma and $DM=11.8$) and a global extinction $A_{V}=2.7$~mag.  \label{cmd_ir}}
    \end{figure}
    
    The corresponding $(J-K_{\textrm{S}})$\,--\,$K_{\textrm{S}}$ diagram is displayed in Fig.~\ref{cmd_ir}. Although differential reddening should be less important in the near infrared, photometric accuracy is worse in the 2MASS catalogue than in \textit{Gaia} DR2. As a consequence, there is still a 0.3~mag spread in $(J-K_{\mathrm{S}})$ at a given magnitude. No object fainter than $K_{\mathrm{S}} = 13.5$ survives our quality cuts, and a few of the matches are very clearly incorrect. Even so, the cluster sequence is well defined. An isochrone with the preferred values from the optical photometry analysis, i.e. $\tau = 75\:\mathrm{Ma}$, $\mu = 11.8$ is shown. The fit is not bad, although we must point out that a value $A_{V}= 2.7$ was used to redden the isochrone. This is somewhat lower than the values used in the fit to \textit{Gaia} data, but the difference is small. It could be pointing to a small deviation from the standard reddening law, but it could also be simply reflecting the inability of the simple polynomial fit of the \citet{cardelli} law to reproduce the effect of high extinction values. A much better characterisation of the extinction law will be possible with the DR3 spectrophotometric data.
   
\section{Discussion}

We have identified a large and populous open cluster in the constellation of Scutum. Analysis of \textit{Gaia} DR2 data confirms its reality and extent\footnote{After the analysis was completed, \textit{Gaia} EDR3 was published. We have checked that the average cluster values for all astrometric parameters in EDR3 are perfectly compatible within the dispersion given with the DR2 data and preferred to retain the analysis based on the likely members of \citet{castrogin20} only}. Its basic parameters are corroborated by an independent analysis by \citet{castrogin20}. Although very strong differential reddening prevents an accurate isochrone fit, its age is well constrained by the \textit{Gaia} DR2 parallax. Ages younger than $70\:\mathrm{Ma}$ or decidedly older than $80\:\mathrm{Ma}$ seem incompatible with the parallax, even after accounting for the possible systematic errors. At a nominal age of $75\:\mathrm{Ma}$, a solar metallicity cluster must have its main-sequence turn-off around B5\,V, in good agreement with the spectral types that we find for the brightest members (Table~\ref{classtab}), although we do not reach deep enough to find any dwarf stars. At this age, the initial mass of the evolved stars is $\approx 6\:\mathrm{M}_{\sun}$, according to the isochrones.

\subsection{Cluster size}
\label{sec:ext}

Valparaiso~1 is a very large cluster on the sky (more than 40 arcmin across, if we consider the region over which both our analysis and that of \citealt{castrogin20} find a non-negligible density of probable members). Given its age, angular size and evolved population, a direct comparison can be made to NGC~6067, one of the most massive clusters of a similar age, with a present day mass of $\approx 5\,000\:\mathrm{M}_{\sun}$ \citep{alonso17}. NGC~6067 has a similar size in the sky, with a tidal radius of $\approx 15$~arcmin at a slightly shorter $DM=11.3$ \citep{alonso17}. Even with the strict selection criteria of \citet{castrogin20}, Valparaiso~1 has 190 members brighter than $G=15$. With the parameters derived for the cluster, this limit corresponds to an absolute magnitude $\approx +0.2$, with the exact value depending on the reddening. This magnitude is typical of a B9\,V star ($\approx3\:\mathrm{M}_{\sun}$). Leaving aside the evolved stars, we thus have about 180 mid- and late-B cluster members. This number can be compared to the $\approx100$ B-type stars in NGC~6067 \citep{alonso17}. NGC~6067 is slightly older and has a higher metallicity than Valparaiso~1. Its main-sequence turn-off is thus found at B6\,V. Therefore the main sequence in Valparaiso~1 should extend a bit more towards early types. Even so, the number of B-type members indicates that Valparaiso~1 is almost certainly more massive than NGC~6067. This is even more evident if we resort to the 2-$\sigma$ membership criterion used in Section~\ref{sec:def}, which gives 257 likely members down to $G=15$. For NGC~6067, \citet{alonso17} estimate an initial mass above $7\,000\:\mathrm{M}_{\sun}$.

Such a high number of members (and hence mass) is borne out by the very large extent of Valparaiso~1. Even accepting the cluster radius from the King profile fit of only 9~arcmin, this implies that the cluster is 12~pc across. In fact, we detect members at distances of 20~arcmin from the nominal centre, equivalent to about 13~pc, as also do \citet{castrogin20}. 

\subsection{CM~Sct}

CM~Sct is a well studied 3.9~d Cepheid, used as part of several period/luminosity calibrations \citep[e.g.][]{tammann03, fouque07}. It has been generally assumed to be a fundamental mode pulsator, and this is confirmed by the recent reassessment by \citet{ripepi19}, using DR2 data. For a fundamental mode pulsator, such a pulsation period is expected at an age around 80~Ma, according to the classical calibration of \citet{bono05}. The accuracy of such aging is debatable, as the models used by \citet{bono05} assume a solar metallicity $Z_{\sun} = 0.02$, while the {\sc parsec} isochrones that we are using take $Z_{\sun} = 0.015$, but the agreement with the cluster age derived from isochrone fitting is very encouraging.

CM~Sct is a halo member of Valparaiso~1, lying about 9.3~arcmin from the centre, close to the edge of the cluster determined by the King profile (which we know to underestimate the true cluster size), but substantially closer than many other members.  About half of the classical Cepheids confirmed as cluster members are halo members of their respective clusters \citep{anderson13}. At 2.3~kpc, this angular distance from the centre corresponds to 6.2~pc, comparable to the distances of V~Cen, EV~Sct or QZ~Nor to their respective cluster centres \citep{anderson13}. CM~Sct lies to the Northeast of the cluster, a region which, as seen in Fig.~\ref{twobranches}, seems to have relatively low extinction. In fact, published photometry for the Cepheid indicates a rather lower reddening than the cluster average.  \citet{fernie90} quotes $E(B-V) = 0.75$, while \citet{laney07} give only $E(B-V) = 0.7$. For a standard reddening law, these values correspond to $A_{V}\approx2.2$\,--\,2.3, similar to the less reddened main sequence members.

Thanks to \textit{Gaia} DR2 astrometric parameters, CM~Sct joins the growing number of classical Cepheids that belong to open clusters. DR2 data have also allowed the identification of the parental cluster of SV~Vul \citep{negueruela20} and associated two previously unknown Cepheids to the open clusters Berkeley~51 and Berkeley~55 \citep{lohr18}. 

\subsection{The nature of star F}

Among the large complement of evolved stars in Valparaiso~1, star~F (TYC~5121-758-1) is remarkable for being much redder and brighter than any other member. In fact, its position on the preferred isochrones suggests it is an early AGB star. As such, it would be the first identification of an AGB star in a young open cluster. Recently, \citet{fragkou19} found a planetary nebula associated to NGC~6067, as discussed above another massive cluster of about the same age. These connections provide for the first time empirical evidence of the evolutionary paths for $\sim6\:\mathrm{M}_{\sun}$ stars. Until now, it had been assumed that the brightest (O-rich) Galactic AGB stars were descended from moderately massive stars ($4$\,--\,$8\:\mathrm{M}_{\sun}$) because of their population characteristics \citep[e.g.][]{gh07}, but direct comparison to models could only be performed for objects in the Magellanic Clouds. The accurate identification of members allowed by \textit{Gaia} for spatially extended clusters, such as NGC~6067 or Valparaiso~1, is starting to give us valuable examples in the Milky Way itself.

Morphologically, star F looks like a normal low-luminosity M-type supergiant. We find no obvious anomalies within the limited spectral range observed. Formally, we classify it as a supergiant (M1\,Ib), because the strength of the luminosity indicators, such as the equivalent width of the \ion{Ca}{ii} triplet or the blend at 8648\,\AA\ \citep[see][and references therein]{dorda16}, put it within this luminosity class. Nevertheless, this classification does not imply any assumption about its mass or internal structure. In fact, its luminosity, $\log \left(L_{*}/L_{\sun} \right) \approx 3.8 $ according to its position on the isochrone, is too low for a real supergiant (i.e. an object with an initial mass above $8\:\mathrm{M}_{\sun}$).

The presence of strong lithium lines, namely at the location of the \ion{Li}{i}~6708\,\AA\ doublet, is believed a telltale sign of a massive AGB star \citep{gh07}. According to modern models \citep[e.g][]{mazzitelli99, vanraai12}, lithium production happens during the early AGB phase and Li surface abundance  reaches  a  maximum  during the early thermal pulse phase. We do not detect a strong line at the position of the \ion{Li}{i}  doublet. We observe a weak feature, typical of stars of the spectral type, which may include a contribution of the Li line, but may also be due to a number of neighbouring metallic lines. Unfortunately, our spectrum lacks the resolution needed for spectral synthesis and an estimate of the Li abundance. In any event, the lack of a strong Li feature does not rule out the early-AGB nature of star F, as its position on the isochrone suggests it has only recently entered the assymptotic branch. An alternative explanation as a red supergiant straggler formed by binary interaction seems ruled out by its low luminosity.
   
\section{Conclusions}
\textit{Gaia} astrometric data reveals that an inconspicuous grouping of mid-B giants is in reality the core of a very large open cluster, which we call Valparaiso~1, comparable in size and mass to the most massive open clusters in the solar neighbourhood. Ten luminous bright giants or supergiants are members of the cluster, including the well-known 4-d Cepheid CM~Sct. Analysis of their intermediate-resolution spectra reveals a solar-like metallicity. Although strong differential reddening renders accurate determination of cluster parameters unfeasible with the current dataset of \textit{Gaia} and 2MASS photometry, the \textit{Gaia} parallax -- implying a distance of $2.3^{+0.7}_{-0.4}\:$kpc, where the errors take into account possible systematic effects --  is only compatible with ages no younger than $\sim70\:\mathrm{Ma}$ and not much older than $\sim80\:\mathrm{Ma}$. Such an age is compatible with the complement of B-type giants observed (in the range B5\,--\,B7\,III) and implies a mass of $\approx6\:\mathrm{M}_{\sun}$ for the more evolved stars.

The similarly aged clusters NGC~6649 and NGC~6664 are pretty close in the sky, lying at about the same distance and also hosting Cepheids. Both clusters contain about 60 B-type stars, implying initial masses somewhat below $3\,000\:\mathrm{M}_{\sun}$ \citep{alonso20}. By comparison, the $>180$ B-type stars in Valparaiso~1 suggest an initial mass approaching $10\,000\:\mathrm{M}_{\sun}$. This number is also borne out by direct comparison to the similarly large NGC~6067 \citep{alonso17}. Both NGC~6649 and NGC~6644 have been studied for decades, while the much larger Valparaiso~1 remained unknown, despite the attention drawn by its Cepheid member, CM~Sct. This suggests that confusion due to high fore- and background contamination has been a major issue preventing the identification of clusters without a strong central concentration. In fact, many of the new open clusters found by \citet{castrogin20} are characterised by low stellar densities and a dispersed structure. In view of this, it is highly likely that substantial numbers of open clusters remain hidden in areas of high stellar density towards the inner Galaxy. As discussed in the Introduction, the age distribution of known massive clusters suggests that some of them may be very massive. 

Unlike in younger clusters \citep[e.g. vdBH~222;][]{marco14}, the bright evolved members of Valparaiso~1 do not stand out at all over their neighbours. Although they are among the brightest stars in the field, there are several dozen stars of comparable (or higher) brightness in the $K_{\mathrm{S}}$ band with similar colours. In view of this, it is highly likely that future searches for clusters based on DR3 and successive releases will find more moderately-young and intermediate-age populous clusters.

Meanwhile, Valparaiso~1 offers us a precious laboratory to study the evolution of intermediate-mass stars, with star~F very likely representing a $6\:\mathrm{M}_{\sun}$ star entering the AGB. Broadband photometry and DR3 spectrophotometry will allow a much more accurate determination of the extinction and hence cluster parameters. 

\section*{Data Availability}
The data underlying this article will be shared on reasonable request to the corresponding author.


\section*{Acknowledgements}

We thank the anonymous referee for helpful comments. This research is partially supported by the Spanish Government under grants AYA2015-68012-C2-2-P and PGC2018-093741-B-C21 (MICIU/AEI/FEDER, UE). HMT is also supported by FCT - Fundação para a Ciência e a Tecnologia through national funds and by FEDER through COMPETE2020 - Programa Operacional Competitividade e Internacionalização by these grants: UID/FIS/04434/2019, UIDB/04434/2020; UIDP/04434/2020, PTDC/FIS-AST/28953/2017, and POCI-01-0145-FEDER-028953. ANC is supported by the international Gemini Observatory, a program of NSF’s NOIRLab, which is managed by the Association of Universities for Research in Astronomy (AURA) under a cooperative agreement with the National Science Foundation, on behalf of the Gemini partnership of Argentina, Brazil, Canada, Chile, the Republic of Korea, and the United States of America. JB and RK are partially funded by ANID – Millennium Science Initiative Program – ICN12\_009 awarded to the Millennium Institute of Astrophysics MAS

The INT is operated on the island of La Palma by the Isaac Newton Group, installed in the Spanish Observatorio del Roque de Los Muchachos of the Instituto de Astrof\'{\i}sica de Canarias. The Starlink software (Currie et al.\ 2014) is currently supported by the East Asian Observatory. This work has made use of data from the European Space Agency (ESA) mission
{\it Gaia} (\url{https://www.cosmos.esa.int/gaia}), processed by the {\it Gaia}
Data Processing and Analysis Consortium (DPAC,
\url{https://www.cosmos.esa.int/web/gaia/dpac/consortium}). Funding for the DPAC
has been provided by national institutions, in particular the institutions
participating in the {\it Gaia} Multilateral Agreement.

 This research has made use of the Simbad, Vizier and Aladin services developed at the Centre de Donn\'ees Astronomiques de Strasbourg, France. It also makes use of data products from 
the Two Micron All Sky Survey, which is a joint project of the University of
Massachusetts and the Infrared Processing and Analysis
Center/California Institute of Technology, funded by the National
Aeronautics and Space Administration and the National Science
Foundation. 



\bibliographystyle{mnras}
\bibliography{clusters,rsgs,bins,gaia,obstars,cepheid,interstellar,analysis} 





\bsp	
\label{lastpage}
\end{document}